\documentclass{article}

\usepackage[english]{babel}

\usepackage[letterpaper,top=2cm,bottom=2cm,left=3cm,right=3cm,marginparwidth=1.75cm]{geometry}

\usepackage{amsmath}
\usepackage{graphicx}
\usepackage{amsthm}
\usepackage{comment}
\usepackage{natbib}
\usepackage{booktabs}
\usepackage{caption}

\usepackage[colorlinks=true, allcolors=blue]{hyperref}

\theoremstyle{definition}

\theoremstyle{assumption}

\title{A simple sensitivity analysis for observational studies}
\author{Jeffrey Zhang\textsuperscript{1}\thanks{jeffzhang@uchicago.edu}, \ Dylan Small\textsuperscript{2}\thanks{dsmall@wharton.upenn.edu}}
\date{
\begin{minipage}{0.95\textwidth}
\centering
\small
\textsuperscript{1}Data Science Institute, University of Chicago \\
\textsuperscript{2}Department of Statistics and Data Science, University of Pennsylvania 
\end{minipage}
}
\begin{document}
\maketitle

\begin{abstract}
Drawing causal conclusions from non-experimental (observational) data is a central aim in science but is extremely challenging. Without control over treatment assignment (as in an experiment), treated and control groups may not be comparable on unmeasured factors affecting treatment and outcome. To address such concerns, sensitivity analyses can be performed; sensitivity analyses acknowledge the existence of unmeasured factors but ask how strong they must be to overturn causal conclusions. We (re)introduce a particularly simple form of sensitivity analysis which can be performed by any reader of an observational study with access to only an estimate and confidence interval for a causal effect. This simple sensitivity analysis may serve as an effective tool for readers wishing to gauge the robustness of conclusions from an observational study.
\end{abstract}

\section{Introduction}
\label{sec: intro}

Assessing cause-and-effect relationships is one of the most important tasks in science. The best device for this task is the randomized controlled trial. Randomization is often either infeasible or unethical, though, so researchers often resort to observational studies to probe causal hypotheses. For example, it would be unethical to withhold a proven life-saving treatment from a patient on the basis of conducting a randomized trial. In a randomized trial, treatment assignment is governed by a haphazard coin flip, resulting in treatment and control groups that are balanced on all causes of both treatment and outcome, observed or not. Consequently, in a randomized trial, observed differences in outcomes between treatment and control group can be solely attributed to the treatment itself. In observational studies, although imbalances in observed causes of treatment and outcome between treatment and control groups can be adjusted for using statistical methods, imbalances in unmeasured ones cannot. This major distinction between randomized trials and observational studies is well-appreciated, and questions of causality have often been deemed unanswerable only using observational data. In fact, in its instructions for authors, the Journal of the American Medical Association prohibits the use of causal language when analyzing non-randomized data \citep{jama_author}. However, in this article, in accordance with \cite{hernan2018c}, we explicitly use causal language so as to not obfuscate the scientific aim, which is to estimate causal effects rather than mere associations. Although one can always logically argue that estimates of causal effects from observational data are biased, it may be possible to assuage concerns about the bias. For example, a \emph{sensitivity analysis} is often advocated to assess the robustness of the causal conclusion to bias from unmeasured factors. In the academic literature, one component of many sensitivity analyses is a parameter that governs how imbalanced the treatment and control groups can be with respect to a hypothetical unmeasured factor that causes both treatment and outcome. 

Sensitivity analysis has a long history in observational studies, dating back to \cite{Cornfield2009}. In recent years, researchers have proposed countless sensitivity analysis methods for various estimands. Most methods are developed from the perspective that the data analyzer/researcher should conduct and report some version of sensitivity analysis. Typically, this requires the analyst to plan for the sensitivity analysis as most cannot be done post-hoc based on point estimates and confidence intervals of treatment effects obtained under the no unmeasured confounding assumption. Few sensitivity analysis frameworks that can accommodate post-hoc conclusions exist. By post-hoc, we mean that a reader of an observational study can quickly assess sensitivity of the results to unmeasured confounding without doing any calculations, and with access only to estimates and confidence intervals, but not the original dataset. A specific sensitivity analysis framework, sometimes termed the bias function approach, originally formalized by Robins, Rotnitzky, and Scharfstein \citep{Robins1999, Robins2000, Scharfstein1999} satisfies this post-hoc criteria, as it is simple, intuitive, and requires no numerical calculations. In fact, it can be described in two steps:
\begin{enumerate}
    \item Suppose that a researcher has obtained a confidence interval, $[l, h]$, for some measure of the difference in outcome between treated and control groups, controlling for observed causes of treatment and outcome. This could be a difference, ratio, or odds ratio.
    \item Suppose the lower confidence bound of the effect estimate is positive (analogous reasoning will apply for the negative case), i.e. $l > 0$, where 0 represents a null effect. For a confidence interval of the effect measure to include 0, there would need to be unmeasured factor(s) that could account for the treated group having outcomes, in the absence of treatment, to be larger than the control groups outcomes by at least $l$.
\end{enumerate}

 We illustrate just how easy the sensitivity analysis is to perform by reexamining several famous successes and failures of observational studies. The ease of doing the sensitivity analysis is a strength, but it can also lead to the analyst being led astray about the robustness of a study's conclusions. This phenomenon is apparent throughout the discussion of the examples. 

\section{Health risk of caffeine}
\cite{MARTIN1987} studied whether caffeine intake of any form (e.g. coffee, tea, drugs) by a pregnant woman affects the birthweight of her baby. They conducted an observational study, controlling for age, marital status, education, race/ethnicity, smoking, alcohol use, marijuana use and number of previous births, and estimated that medium caffeine intake by a mother (defined as 150--300 mg per day) increased the risk of a low birth weight baby by 2.3 times with a 95\% confidence interval of (1.1, 5.2) times increased risk compared to mothers who did not consume caffeine. We now apply the sensitivity analysis outlined in the previous section. The lower bound of the confidence interval is 1.1, so there would need to be unmeasured factors (after accounting for age, marital status, education, etc.) other than caffeine that make the mothers drinking 2-3 cups of coffee a day 1.1 times more likely to have a low birth weight baby for us to no longer be confident that a pregnant woman drinking 2-3 cups of coffee a day reduces the birthweight of her baby. Although important confounding factors such as smoking and alcohol use were controlled for, more fine-grained factors were not. For example, a hypothetical skeptical reader could argue that though marijuana use was measured, cocaine use during pregnancy was not, and a mother who used cocaine may be more likely to drink coffee and have a low birth weight baby. The sensitivity analysis reveals that even a small amount of confounding from such unmeasured factors (factors besides caffeine that make the mothers drinking 2-3 cups of coffee a day during pregnancy 1.1 times more likely than the matched non-coffee drinkers) could nullify the evidence that drinking 2-3 cups of coffee causes low birth weight. Indeed, later on, a randomized trial that investigated a distinct but highly related question revealed that the association found in the observational study was not causal. In the trial, women who were drinking at least three cups of coffee at the beginning of their pregnancy were randomized to either replace their usual coffee with decaffeinated coffee or to just continue drinking their usual coffee. \cite{Bech2007} found no effect of drinking the decaffeinated coffee on low birth weight. This example demonstrates that typically, study findings with small effects are easily overturned by small amounts of unmeasured confounding. Thus, such findings should be heavily scrutinized. At the same time, a small effect magnitude is not the only criteria for judging the robustness of a causal finding, which we illustrate in the ensuing example. 

\section{Long term effects of losing a spouse or child in a motor vehicle crash}
In observational studies, a statistical association may appear small, but the absolute magnitude is not what is most important, but rather the magnitude relative to plausible biases. Plausible biases can differ widely from study to study, and they mainly hinge on how ``random'' a treatment assignment is. In the best circumstance, researchers can find situations where nature haphazardly assigns people to treatment or control, much like in a randomized trial. Such scenarios are often referred to as natural experiments. One example comes from a study conducted by Darrin Lehman, Camille Wortman, and Allan Williams. They examined whether the loss of a spouse or a child in a motor vehicle crash 4 to 7 years prior causes depression in the present \citep{Lehman1987}. \cite{Lehman1987} restricted their analyses to spouses who were married and living together and children less than 18 years of age at the time of the crash. Drivers involved in car accidents in which they were deemed not responsible are not expected to be systematically different than the general population of drivers on factors related to depression, such as psychological disorders. In contrast, drivers that were deemed responsible for accidents may have a higher rate of psychological disorders related to depression. Thus, \cite{Lehman1987} restricted their analysis to drivers deemed not responsible. In so doing, their analysis is less likely to be biased by important unmeasured factors strongly related to later-life depression. They controlled for observed factors by matching individuals who lost a spouse or child in a crash to married controls who came to renew a driver's license on the basis of gender, age, family income, education, number and ages of children. Their study found that of the individuals who had lost a loved one, 31\% ``could not shake off the blues even with help from their family and friends'' on at least 3 of the past 7 days compared to just 10\% for the matched controls. The 95\% confidence interval for the percentage point difference between the bereaved individuals and matched controls is (6\% , 41\%). A simple application of the sensitivity analysis states that for there to be no statistical evidence of a causal effect of losing a loved one to a car crash on depression 4 to 7 years later, the people who had lost a loved one would have had to be 6 percentage points more likely to be depressed than their controls, due to factors besides losing a loved one and the measured factors. At first glance, 6 percentage points does not seem to be a huge difference, and may appear roughly comparable to the size of the association from the coffee and birthweight example. Moreover, there is always the possibility of unmeasured differences between those that lost a loved one and the controls, such as frequency of driving. At the same time, given that getting into a car crash that one is not responsible for is a relatively random occurrence, 6 percentage points seems more unlikely than the 1.1 times increased risk from the coffee and birthweight example. We have now seen that natural experiments can be powerful tools for drawing robust causal conclusions, but in practice they are quite rare. Fortunately, they are not the only means to establishing causality.

\section{Smoking and lung cancer}
Perhaps the greatest success story of observational studies is the discovery of smoking as a cause of lung cancer. For obvious ethical reasons, no randomized controlled trials have been conducted that assign smoking, but there is no doubt about the causal link. How are we so sure? Some of the first epidemiological evidence that suggested a causal link arose in the 1950s and early 1960s \citep{Hammond1958, Haenszel1958, Dorn1959, HAMMOND1964}. For example, \cite{HAMMOND1964} conducted a prospective study initiated in 1959 that matched 36,975 cigarette smokers (men currently smoking 20 or
more cigarettes a day at enrollment) to non-smokers (men who never smoked regularly) on the basis of age, height, race, nativity, religion, marital status, residence (urban or rural), certain occupational exposures, education, drinking habits, nervous tension, use of tranquilizers, sleep, exercise, health at time of enrollment, and past history of cancer, heart disease, stroke, and high blood pressure. Hammond found a mortality ratio of 9.17 when comparing lung cancer deaths between cigarette smokers and non-smokers, with an associated 95\% confidence interval of (5.04, 18.28). Findings such as these were met with a great deal of backlash. R.A. Fisher, a prominent statistician, among other critiques, posited the existence of an unmeasured genetic factor that causes both smoking and lung cancer \citep{Fisher1958}. In response to this critique, \cite{Cornfield2009} conducted what is regarded as the first sensitivity analysis. In fact, the sensitivity analysis they conducted is exactly of the form of sensitivity analysis described in Section \ref{sec: intro}.  \cite{Cornfield2009} acknowledged that such a genetic factor might exist, but it would have had to have been ``at least nine-fold more prevalent among cigarette smokers than among non-smokers" in order to explain away the observed association between smoking and lung cancer, which they deemed scientifically implausible. Based on the study of \cite{HAMMOND1964}, if one wished to also account for sampling uncertainty, the gene would still need to be at least five-fold more prevalent. In this example, the magnitude of the observed association was so large that it was difficult to imagine an unmeasured factor that could have overturned the qualitative conclusion, especially the genetic factor under consideration. Interestingly, \cite{VanderWeele2012} found that variants on chromosome 15q25.1 affect smoking behavior and also directly affect lung cancer, supporting Fisher's hypothesis of the existence of an unmeasured genetic factor that causes both smoking and lung cancer. However, the odds ratio for the potential outcome of lung cancer for a fixed smoking level with two alleles compared to zero alleles of the variant was 1.7 (95\% confidence interval: 1.3, 2.2), which is not nearly enough to explain away the observed association between smoking and lung cancer. Through the first three examples, we have now seen cases where a small observed association was non-causal, a small observed association was causal, and a large observed association was causal. Though it is typically true that larger associations are more robust to unmeasured confounding and thus more likely to be causal, this is not always the case.

\section{Vitamin C and cancer}
In contrast to the smoking and lung cancer success, a great failure of observational studies was the erroneous ``discovery" of vitamin C as a treatment for cancer. In fact, it was two-time Nobel laureate Linus Pauling that was deceived by the large observed association between vitamin C intake and decreased cancer risk. \cite{Cameron1976} compared 100 terminal cancer patients who were given 10 grams of supplemental Vitamin C per day as part of their treatment i.e., non-randomly prescribed vitamin C, to 1000 similar patients (on the basis of age, gender, cancer location, tumor type, treating clinician) who were treated the same but with no vitamin C. They found that the mean survival time was roughly 5 months greater for those that took vitamin C (about 210 days) than for the controls (about 50 days). More precisely, a point estimate for the average survival time difference was 159.8 days (95\% CI, 99.8 to 219.8 days). In this example, the magnitude of the observed association appears to be quite large, just as in the smoking and lung cancer example. Applying the sensitivity analysis from Section \ref{sec: intro}, for the association to be explained away, there would need to be unmeasured factors after accounting for age, gender, cancer location, and tumor type that made patients taking vitamin C live 100 days longer. It is tempting to conclude that it is unlikely for such unmeasured factors to exist, and therefore there is a causal link. In fact, Pauling believed in the causal link so much that he began taking large quantities of vitamin C. However, the magnitude of the association should always be evaluated with respect to the factors that were actually measured. Important factors that influence vitamin C consumption are likely to be highly associated with survival and were omitted included stage of disease and sites of metastasis. William D. DeWys, chief of the clinical investigations
branch of the National Cancer Institute’s cancer therapy
program, raised these concerns about the comparability of the treated and control groups. Moreover, he found evidence that the treatment group was deemed terminally ill much earlier than the control group. Later on, several randomized controlled trials confirmed DeWys's suspicions, finding no effect of vitamin C on survival in patients with advanced cancer \citep{Moertel1985}. This example demonstrates the importance of gauging the strength of association with respect to the observed factors already adjusted for. Without adjusting for important confounding factors, ``large" associations should not be regarded as strong evidence for causality.

\section{Calibrating a sensitivity analysis: the effect of a sibling death on years of education}
In the previous examples, we have only discussed how a reader could reason about the \emph{absolute} strength of unmeasured factors that could overturn a causal conclusion. Since this may not always be possible, we also discuss an approach to reason about the \emph{relative} (to measured factors) strength of unmeasured factors that could overturn a causal conclusion. To ground ideas, we introduce a study by \cite{Fletcher2013}, who, using data from the National Longitudinal Study of Adolescent Health (Add Health), compared the years of education obtained by sisters who had a sibling die during formative years with those that did not. Employing a regression analysis (see Table \ref{tab: reg analysis}), they adjusted for important measured factors including the surviving sister's age, race, adoption status, birth order, whether they are Catholic, characteristics of the mother, and parental education. The analysis found that a sibling death had an average effect of -0.52 years of education (95 \% CI, -0.70 to -0.34 years). To deem this association non-causal, there would need to be unmeasured factors other than age, race, adoption status, birth order, whether they are Catholic, characteristics of the mother, and parental education that would have made sisters who lost a sibling pursue 0.34 less years of education on average (compared to sisters who didn't lose a sibling) even if counterfactually they hadn't lost a sibling. In this particular example, it may be difficult to determine whether this minimum strength of unmeasured factor needed to overturn the qualitative conclusion is large or small. However, especially in cases where one is worried about a specific unmeasured factor, it may be possible to compare this minimum strength to the strength of a measured factor, which is estimable. We call this \emph{calibrating} the sensitivity analysis, and describe and apply it to the sibling death example. 
\begin{table}[ht]
\centering
\caption{Summary of regression analysis from Table 3 of \cite{Fletcher2013} assessing the effect of sibling death on education of sisters.}
\begin{tabular}{lcc}
\toprule
& point estimate & (SE) \\
\midrule
Death of Sibling & $-0.520^{***}$ & (0.090) \\
Age & $0.046^{**}$ & (0.018) \\
Black & $-0.054$ & (0.113) \\
Hispanic & $0.107$ & (0.089) \\
Adopted & $-0.433^{***}$ & (0.165)  \\
Birth Order & $-0.077^{***}$ & (0.026) \\
Catholic & $0.188^{**}$ & (0.076) \\
Maternal Age at Child’s Birth & $0.048^{***}$ & (0.006)  \\
Mother Had Teen Birth & $-0.232^{***}$ & (0.077) \\
Maternal Education & $0.183^{***}$ & (0.025) \\
Paternal Education & $0.186^{***}$ & (0.018)\\
\midrule
 \textit{*$p < .05$; **$p < .01$; ***$p < .001$} \\
\bottomrule
\end{tabular}
\label{tab: reg analysis}
\end{table}

\subsection{A calibration procedure}
In examples such as this one, where the outcome is continuous and a standard linear regression analysis is performed, it is relatively straightforward to calibrate the sensitivity analysis to a measured factor. One limitation of the sibling death study is that characteristics of the deceased sibling were not recorded and thus could not be adjusted for \citep{Fletcher2013}. A potentially important binary unmeasured characteristic is whether the deceased sibling was disabled, as this could impact whether they die at an early age as well as their sibling's educational attainment. The effect of this unmeasured factor could be calibrated against the effect of whether the mother had a teen birth.  Let $\tau$ denote the coefficient in front of the treatment variable sibling death (representing the causal effect) when running a standard linear regression.  We might believe that the effect of the deceased sibling being disabled (the unmeasured factor) on the surviving sibling's educational attainment does not exceed that of whether the mother had a teen birth (a measured factor). 

\begin{figure}
    \centering
    \includegraphics[width=0.8\linewidth]{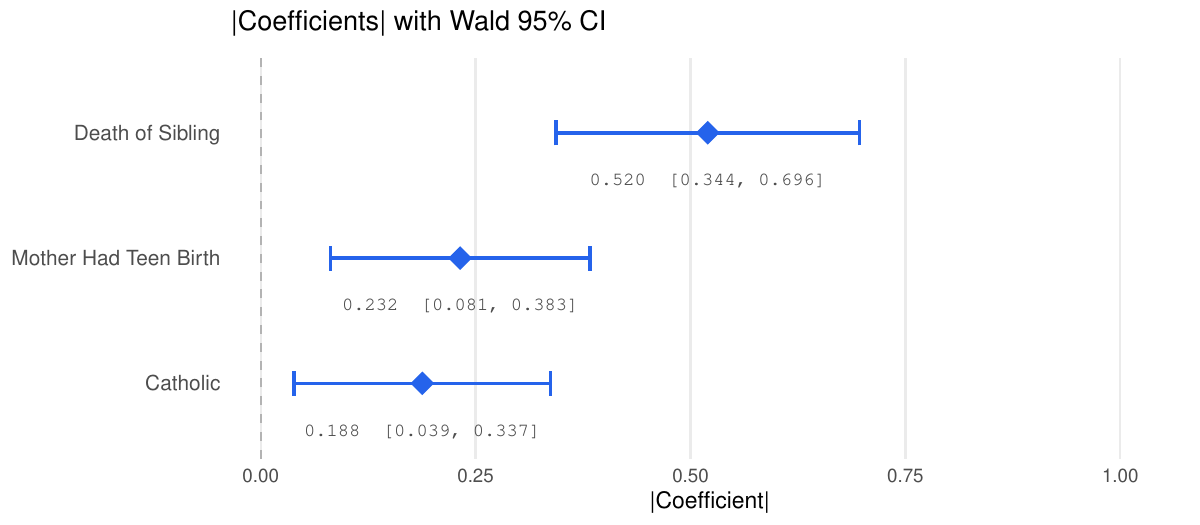}
    \caption{Absolute point estimates and confidence intervals for `Death of Sibling' (treatment of interest), `Mother Had Teen Birth' and `Catholic' (measured covariates) from the regression analysis. Under the assumption that having a disabled deceased sibling (unmeasured factor of concern) has less influence on sibling educational attainment (outcome of interest) than being Catholic, there is very strong evidence that the effect of sibling death is negative, since the confidence interval for the magnitude of the `Death of Sibling' coefficient lies fully to the right to that of `Catholic'. The interval for `Death of Sibling' mostly (but not fully) lies to the right of the interval for `Mother Had Teen Birth' overlap slightly, though the magnitude of the point estimate of the former exceeds that of the latter.}
    \label{fig: calibration plot}
\end{figure}

It then follows that if the magnitude of the coefficient corresponding to teen birth from running a standard linear regression is less than $|\tau|$, the true causal effect must be the same sign as the estimated effect $\tau$ (ignoring sampling variability). In the sibling death study, the coefficient for teen birth from a standard linear regression \citep{Fletcher2013} was estimated to be -0.232. $\lvert-0.232\rvert$ is indeed smaller than $\lvert-0.52\rvert$ (-0.52 was the point estimate for the coefficient for sibling death), lending strong evidence for a causal effect if one believes that the effect of the deceased sibling being disabled on sibling's educational attainment does not exceed that of the mother having a teen birth (and calibrating based on the point estimates). Calibrating based on the confidence limits, however, would yield inconclusive results, since the lower limit for the coefficient for teen birth is $-0.383$, which has slightly larger magnitude than the upper limit for the coefficient for sibling death ($-0.344$). For a visual depiction and an analogous exercise utilizing whether the family is Catholic as the measured factor to calibrate against, see Figure \ref{fig: calibration plot}. Generally, this motivates a simple procedure for a reader of an observational study to gauge the likelihood that a failure to account for an unmeasured factor could render an observed association null. The reader simply needs to compare the estimated coefficient of the treated variable in the regression analysis to the coefficient for the particular measured variable for which they believe has a stronger influence than a hypothetical unmeasured factor. If the magnitude of the coefficient for the treatment is larger than that of the particular measured factor, then the reader should feel confident that the association is indeed causal. This type of exercise can be used to supplement the simple sensitivity analysis from Section \ref{sec: intro}. In this way, the calibration procedure can be a useful tool when the influence of the unmeasured factor on the outcome does not exceed that of a measured factor. We do note that when a more complicated analysis is performed, such as a regression with interactions, the above procedure is not appropriate. However, it is broadly applicable since in many studies, the primary analysis is a standard linear regression analysis without interactions.

\section{Discussion}
Though drawing causal inferences from observational data is a challenging task, tools like the sensitivity analysis from Section \ref{sec: intro} can help. As we saw in the four examples, the sensitivity analysis can be used by a reader to judge how convincing the evidence for a causal link is. The sensitivity analysis states that for an observed association to be deemed non-causal, there must be an unmeasured factor that can completely explain away the association between treatment and outcome. A skeptical reader would then need to posit a sufficiently strong unmeasured factor. At the same time, all conclusions drawn from a sensitivity analysis must be placed in context. Non-skeptical readers must remember that observed associations typically cannot be attributed to a causal effect when important causes of treatment and outcome are unmeasured or unaccounted for, no matter how large the association is.  Overall, the sensitivity analysis from Section \ref{sec: intro} is simple to apply, requiring no time, resources, or statistical expertise. We view it as a useful tool to probe whether an observed association is indeed causal.

\bibliographystyle{plainnat}
\bibliography{references}
\end{document}